\documentclass[
    ,final            
  ]
  {aipproc}

\layoutstyle{8x11single}

\newcommand{\apj}{{Astrophys.\ J.}}
\newcommand{\apjl}{{Astrophys.\ J.\ Let.}}
\newcommand{\aspc}{{ASP Conf.\ Series}}
\newcommand{\mnras}{{Mon.\ Not.\ R.\ Astron.\ Soc.}}

\newcommand{\chandra}{{\it Chandra}}

\newcommand{\msol}{$M_{\odot}$}

\begin{document}

\title{Modeling SN~1996cr's X-ray lines at high-resolution: \\
\it{Sleuthing the ejecta/CSM geometry} }




\author{Daniel Dewey}{
  address={MIT Kavli Institute for Astrophysics and Space Research, Cambridge, MA 02139, USA}
}

\author{Franz E. Bauer}{
  address={Dept.\ de Astronom\'{\i}a y Astrof\'{\i}sica, Pontificia
	   U.\ Cat\'{o}lica de Chile, Casilla 306, Santiago 22, Chile}
}

\author{Vikram V. Dwarkadas}{
  address={Dept.\ of Astronomy and Astrophysics, U.\ of Chicago, 
  5640 South Ellis Avenue, Chicago, IL 60637, USA}
}

\begin{abstract}
SN~1996cr, located in the Circinus Galaxy (3.7 Mpc, $z\sim 0.001$) was
non-detected in X-rays at $\sim$\,1000 days yet brightened to
$L_x \sim 4\times 10^{39}$ erg/s (0.5-8 keV) after 10 years (Bauer et al. 2008).
A 1-D hydrodynamic model of the ejecta-CSM interaction produces good agreement
with the measured X-ray light curves and spectra at multiple epochs.
We conclude that the progenitor of SN~1996cr could have been a massive star,
$M > 30$~\msol,
which went from an RSG to a brief W-R phase before
exploding within its $r\sim 0.04$ pc wind-blown shell (Dwarkadas et
al. 2010). Further analysis of the deep \chandra\ HETG observations
allows line-shape fitting of a handful of bright Si and
Fe lines in the spectrum. The line shapes are well fit by axisymmetric
emission models with an axis orientation $\sim$\,55 degrees to our
line-of-sight. In the deep 2009 epoch the higher
ionization Fe XXVI emission is constrained to high lattitudes:
the Occam-est way to get the Fe H-like
emission coming from high latitude/polar regions is to have more CSM
at/around the poles than at mid and lower lattitudes, along
with a symmetric ejecta explosion/distribution.
Similar CSM/ejecta characterization may be possible for other SNe and, with
higher-throughput X-ray observations, for gamma-ray burst remnants as well.
\end{abstract}

\maketitle


\section{Introduction to SN~1996cr}

\begin{figure}
  \includegraphics[width=0.70\textwidth]{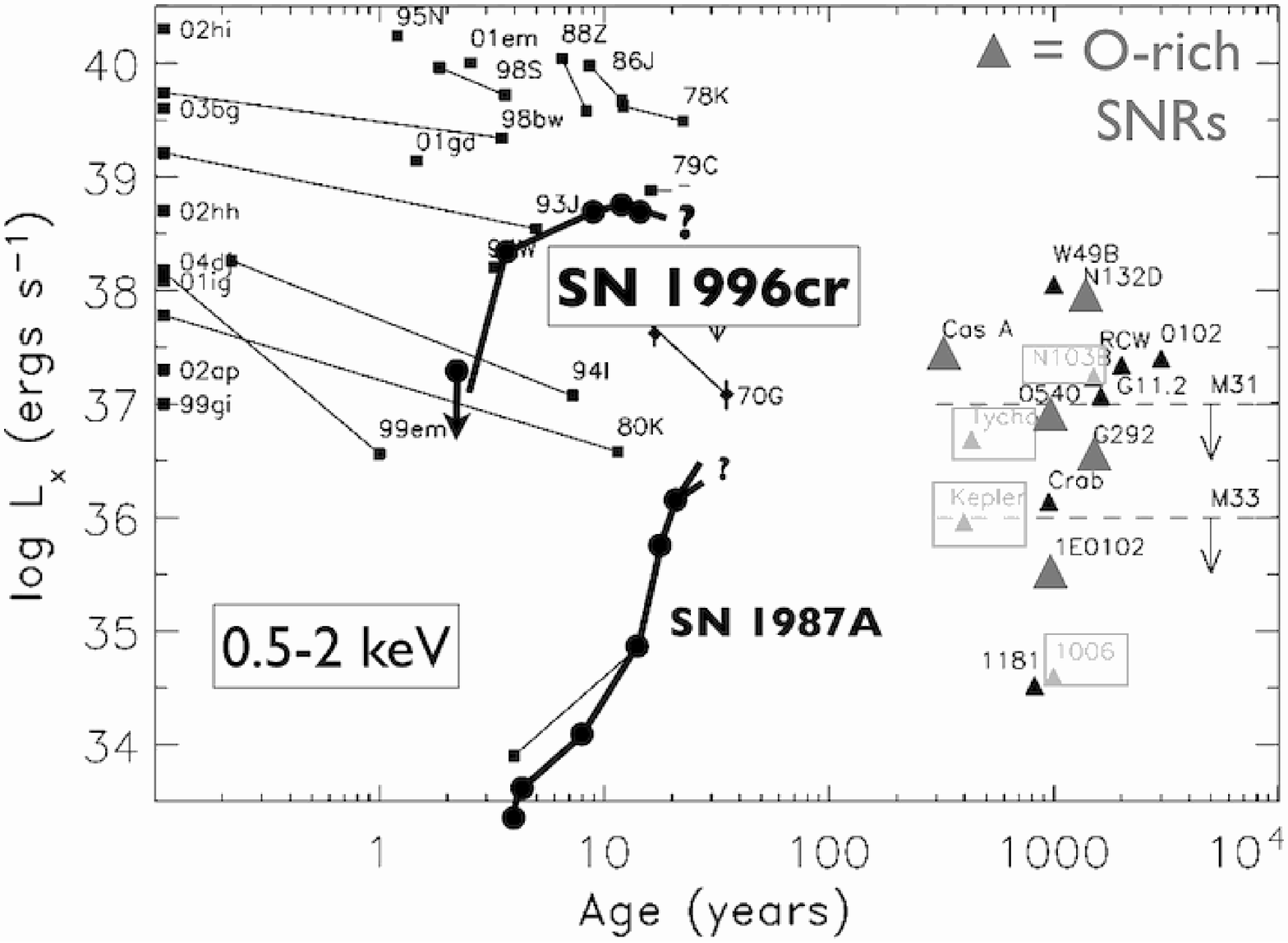}
  \caption{The unique behavior of SN~1996cr and SN 1987A.  Their X-ray light
  curves are overplotted on an $L_X$ vs Age plot
  from \citet{Immler05}.  Unlike other X-ray-bright SNe, these two
  show a dramatic (re-)brightening at ages of a few years to decades.
  As these and other SNe age we will begin to fill in the CC SNe-SNR gap
  seen between ages of 30 to 300 years.
  \label{fig:lightcurve}}
\end{figure}

SN~1996cr, located comparatively nearby in the Circinus Galaxy (3.7 Mpc, $z\sim 0.001$)
was serendipitously detected as a ULX at an age of $\sim$\,5 years \citep{Bauer01}.
As detailed in \citet{Bauer08}\,:~
i) followup archival research and VLT observations (at age $\sim$\,10 years)
allowed SN~1996cr to be classified as a Type IIn SNe, and
ii) archival X-ray data showed that it was not detected at an
age of $\sim$\,1000 days, yet brightened to $L_x \sim 4\times 10^{39}$ erg/s
(0.5-8 keV) after 10 years.  This X-ray behavior is
shared only with SN 1987A, and is roughly shown in Figure~\ref{fig:lightcurve}
in the context of other X-ray-detected SNe and core-collapse (CC) SNRs.
Because of this behavior, and given
an indication of Doppler structure in 2004 \chandra\ HETG data,
we proposed and obtained a deep, 485~ks HETG observation (PI Bauer)
near the beginning of 2009.  
In large part because of the high quality
of this recent data we were able to tune a 1-D hydrodynamic model to agree with the
{\it multi-epoch} X-ray data (next section), and we are now
investigating signatures of the ejecta/CSM geometry in the HETG line shapes (last section.)

\section{Hydrodynamic Model and X-ray Emission}

\begin{figure}
  \includegraphics[width=0.70\textwidth]{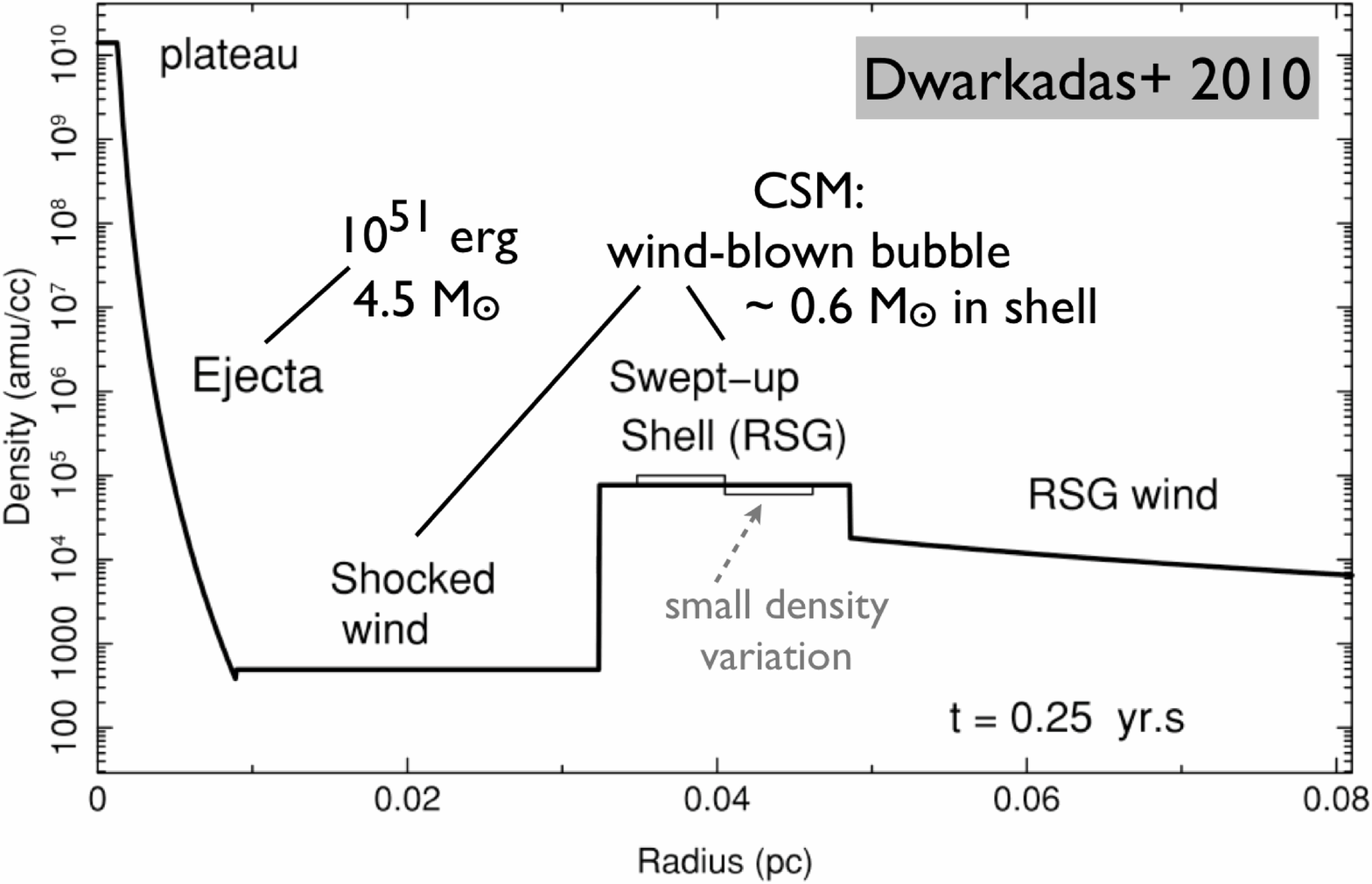}
  \caption{Initial density profile for the hydrodynamic model of SN~1996cr.
  Several months after the SN explosion the ejecta (near the origin at left)
  is seen making its way into the low density W-R wind cavity while
  the shell of swept-up RSG wind material doesn't know what's
  going to hit it.  See Figure~3 of \citet{Dwarkadas10} for snapshots
  of the further evolution of the hydrodynamics beyond this initial
  configuration.
  \label{fig:1dhydro}}
\end{figure}

A 1-D (spherical) hydrodynamic model of the ejecta-CSM interaction
is developed in \citet{Dwarkadas10}; post-processing calculates the X-ray
emission from the non-radiative shocks, showing good agreement
with the measured X-ray light curve and spectra at multiple epochs.
Given our inferred configuration, Figure~\ref{fig:1dhydro},
a realistic evolutionary scenario for SN~1996cr's progenitor has it evolving
from the RSG to the W-R stage, creating a wind-blown bubble with a dense
shell at about 0.04~pc, and then exploding as a SN.
Some conclusions from the modeling are that:
\begin{itemize}
\item
  The 1-D model explains the majority of the observed X-ray continuum
  and lines, and their variation in time.
\item
  The velocities of plasma in the model agree with the scale of the line
  broadening seen in the HETG data.
\item
  The inner ejecta core is opaque to HETG X-rays as late as 2009
  when it has a plateau density of $10^5$~amu\,cm$^{-3}$ and a
  radius of 0.065~pc, giving a column density along the 
  diameter of 4.0\,$\times$\,$10^{22}$~amu\,cm$^{-2}$ of high-Z material.
\item
  Initially the flux from the forward-shocked shell dominates.  However,
  at $\sim$\,8 years the reverse-shocked ejecta contributes 50\% of the flux, 
  this fraction grows to $\sim$\,70\% at 15--20 years.
\item
  Some small fraction (by mass/volume) of denser CSM is needed to produce
  some low-ionization lines seen in the X-ray spectrum; this ``clump''
  emission is a small perturbation on the main hydrodynamics.
\end{itemize}

\section{\chandra\,/\,HETG Line Shapes and Ejecta-CSM Geometry}

\begin{figure}
  \includegraphics[width=0.50\textwidth]{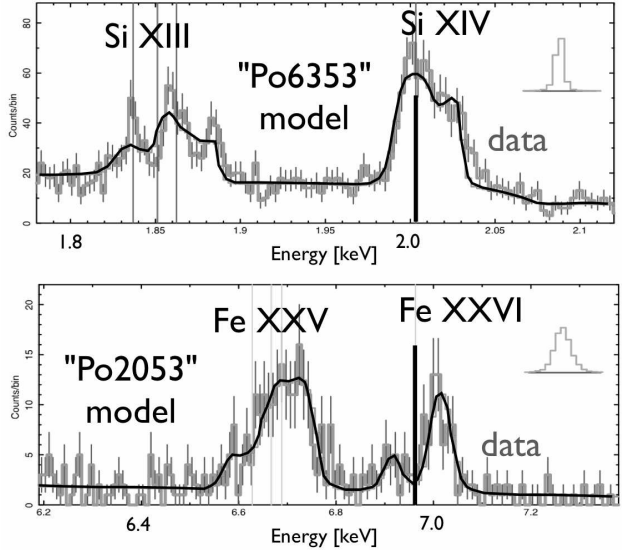}
  \caption{Line shapes of the bright Si and Fe-K lines in SN~1996cr.
  Lines of Si (top) and Fe (bottom) show Doppler-modified structure
  with velocity components of several thousand km/s; the HETG
  instrumental profile is shown in the upper right.  Note the very
  different shapes of the H-like Si XIV and Fe XXVI lines with respect
  to their nominal line energies (vertical black lines.)  The solid black
  curves are model fits based on our 1-D hydrodynamic model velocities
  and an emission region that does not cover a full sphere, see
  Figures~\ref{fig:lineshapes}~\&~\ref{fig:contours}.
  \label{fig:linefits}}
\end{figure}

With a general hydrodynamic picture established, we are now
looking at detailed line-shape fitting of a handful of the lines,
in particular Si and Fe-K lines shown in Figure~\ref{fig:linefits}.
Using simple 3-D modeling techniques \citep{Dewey09} we compared
the HETG data with the line shapes expected from our 1-D model (with spherical symmetry
and an opaque core) and found that, while there is qualitative agreement, there are
statistically significant departures from this simple model.
Since the SNe is unresolved in the X-ray (and only just resolved
by VLBI in 2007 \citep{Bauer08}) we explore
simple, plausible geometric modifications that can better fit the line shapes.

\begin{figure}
  \includegraphics[width=0.75\textwidth]{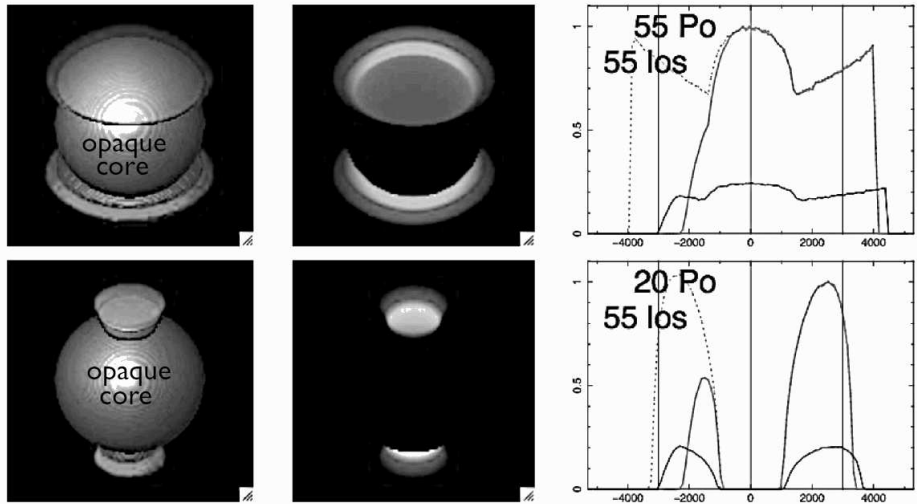}
  \caption{From geometry to line shape.  We calculate the line shape from
  emission that is confined within a range of the pole, e.g., within 55
  degrees (upper row) and within 20 degrees (lower row.)  Including an
  opaque core and viewing the system at an angle to the line-of-sight,
  here 55 degrees, produces a variety of posible line shapes (right
  plots.)  The middle images would be seen by a high spatial-resolution
  X-ray telescope.
  \label{fig:lineshapes}}
\end{figure}

One option is to retain the 1-D radial solution (i.e., densities and
velocities), but to restrict
the interaction and emission to less than the full
$4\pi$ solid angle; examples of such geometries are shown in Figure~\ref{fig:lineshapes}.
In reality, this approximation could be appropriate if either the CSM is
non-uniform or if the explosion/ejecta is not spherically symmetric.
Of course this truncation of the solid angle also reduces the model flux, but
a flux factor of up to a few can be accomodated with
small changes to the parameters (density, radii) and still
obtain light curve and spectral agreement.

\begin{figure}
  \includegraphics[width=0.48\textwidth]{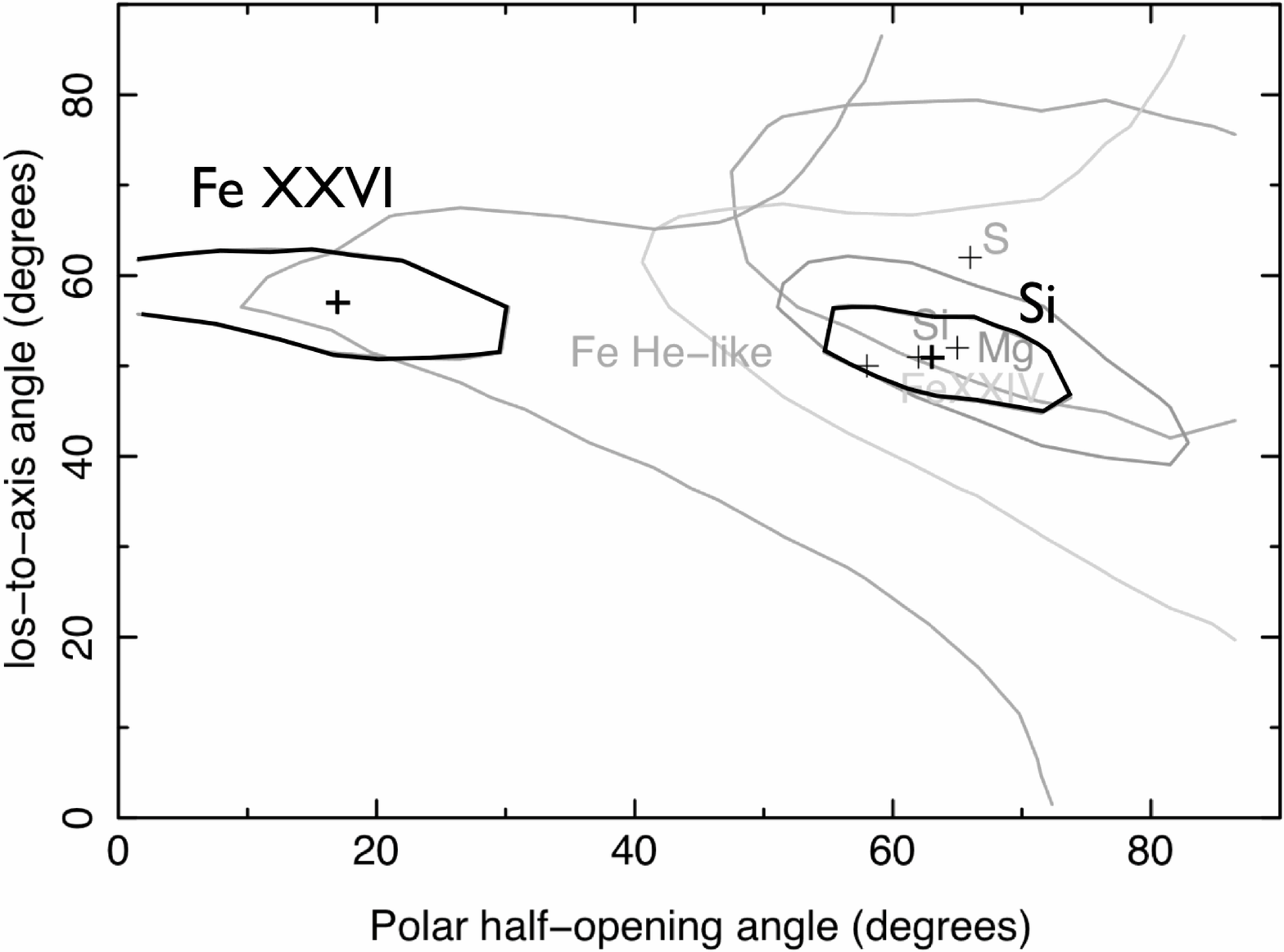}
  \includegraphics[width=0.31\textwidth]{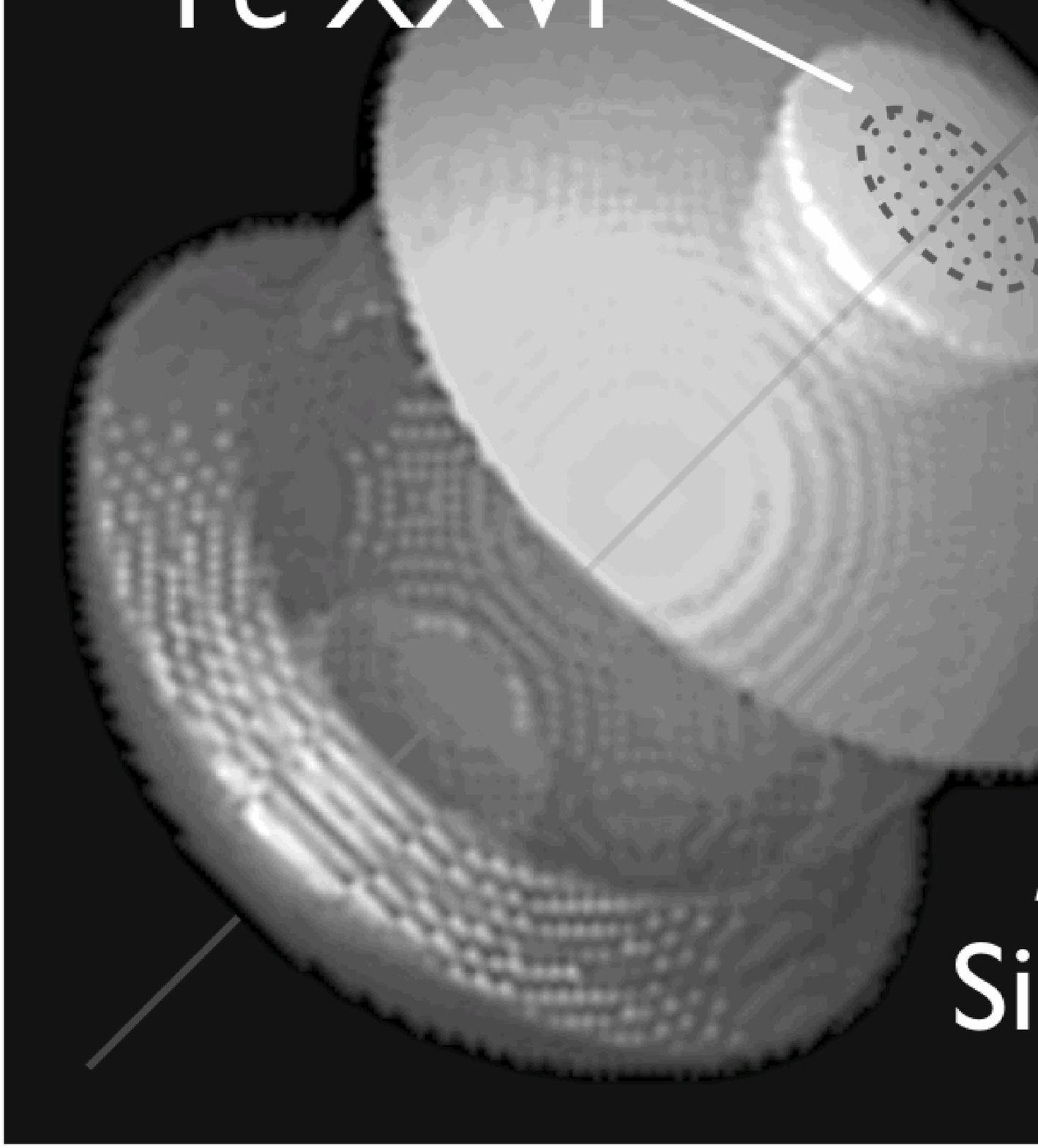}
  \caption{Fitting the emission-line geometries.  Using line
  shapes specified
  by a polar angle and the angle to the line-of-sight, we
  generate 1-sigma confidence contours for fits to discrete lines, left.  The Si,
  S, Mg, and the lower-ionization Fe XXIV lines are all fit with similar
  parameters.  In contrast the Fe XXVI line is much better fit with its
  emission confined to higher-lattitudes.  The solid black lines in 
  Figure~\ref{fig:linefits} show the Si and Fe fits; the image at
  right is the implied geometry.
  \label{fig:contours}}
\end{figure}

Comparing models with data in Figures~\ref{fig:linefits}~\&~\ref{fig:contours},
we find that the line shapes are well fit by ``polar''
models consisting of uniform emission within a half-opening angle of
the poles: within $\sim$\,60--70 degrees for Si and other lines
and within $\sim$\,10--30 degrees for
the Fe XXVI line. In all cases the polar axis is consistent with being
oriented at $\sim$\,50--60 degrees to our line-of-sight; this leads to the
schematic image shown in Figure~\ref{fig:contours} (the position angle
is arbitrary.)

A preliminary sensitivity study of the effects of changing the
hydrodynamic parameters
suggests that the easiest way to get more Fe XXVI emission is to
increase the swept up mass at high lattitudes, rather than, say,
changing the ejecta mass or energy along that direction. 
We are currently in the processes of creating a more realistic model to take this
and the optical line shapes into account.  Perhaps in the future
similar CSM/ejecta sleuthing will be possible for more SNe
and even gamma-ray burst remnants.

\begin{theacknowledgments}
DD was supported by NASA through SAO contract SV3-73016 to MIT for
Support of the \chandra\ X-Ray Center (CXC) and Science Instruments.
FEB and VVD were supported by NASA through the \chandra\ Guest
Observing program, Award Numbers GO9-0086A/B and GO0-11095A/B issued by
the CXC. The CXC is operated by the SAO for and on
behalf of NASA under contract NAS8-03060.
\end{theacknowledgments}



\bibliographystyle{aipproc}   

\begin{thebibliography}{9}


\bibitem[\protect\citeauthoryear{Bauer et al.}{2001}]{Bauer01}
Bauer, F.E., et al. \apj\ \textbf{122}, 182 (2001).

\bibitem[\protect\citeauthoryear{Bauer et al.}{2008}]{Bauer08}
Bauer, F.E., et al. \apj\ \textbf{688}, 1210 (2008).

\bibitem[\protect\citeauthoryear{Immler \& Kuntz}{2005}]{Immler05}
Immler, S.\ \& Kuntz, K.D. \apjl\ \textbf{632}, L99 (2005).

\bibitem[\protect\citeauthoryear{Dwarkadas et al.}{2010}]{Dwarkadas10}
Dwarkadas, V.V., Dewey, D.\ \& Bauer, F. \mnras\ \textbf{407}, 812 (2010).

\bibitem[\protect\citeauthoryear{Dewey \& Noble}{2009}]{Dewey09}
Dewey, D.\ \& Noble, M.S.  ADASS XVIII, \aspc\ \textbf{411}, 234 (2009).

\end{thebibliography}


\end{document}